\documentclass[10pt, conference, letterpaper]{IEEEtran}
\usepackage{cite}
\usepackage{amsmath,amssymb,amsfonts}
\usepackage{algorithmic}
\usepackage{graphicx}
\usepackage{textcomp}
\usepackage{xcolor}
\def\BibTeX{{\rm B\kern-.05em{\sc i\kern-.025em b}\kern-.08em
    T\kern-.1667em\lower.7ex\hbox{E}\kern-.125emX}}

\usepackage{preamble/orcidlink}
\usepackage[nolist,nohyperlinks]{acronym}
\usepackage[detect-weight=true, detect-family=true]{siunitx}
\DeclareSIUnit{\belmilliwatt}{Bm}
\DeclareSIUnit{\bel}{B}
\DeclareSIUnit{\bitpersecond}{bps}
\DeclareSIUnit{\samplepersecond}{Sps}
\usepackage{placeins}
\usepackage[inkscapelatex=false]{svg}
\usepackage{multirow}
\usepackage{enumerate}
\usepackage[inline]{enumitem}

\makeatletter
\newcommand{\linebreakand}{%
  \end{@IEEEauthorhalign}
  \hfill\mbox{}\par
  \mbox{}\hfill\begin{@IEEEauthorhalign}
}
\makeatother
\begin{acronym}
    \acro{IoT}{Internet of Things}
    \acro{BLE}{Bluetooth Low Energy}
    \acro{RSSI}{received signal strength indicator}
    \acro{MCU}{microcontroller unit}
    \acro{SoC}{system on a chip}
    \acro{HAL}{Hardware Abstraction Layer}
    \acro{SDK}{Software Development Kit}
    \acro{FDMA}{frequency division multiple access}
    \acro{TDMA}{time division multiple access}
    \acro{pps}{packets per second}
    \acro{TTL}{time to live}
    \acro{SMA}{SubMiniature version A}
    \acro{EMI}{electromagnetic interference}
    \acro{IQR}{Interquartile Range}
    \acro{GATT}{Generic Attribute Profile}
    \acro{GAP}{Generic Access Profile}
    \acro{LL}{Low-Latency}
    \acro{LE}{Low-Energy}
\end{acronym}

\usepackage{comment}

\usepackage{eso-pic}
\usepackage{url}
\AddToShipoutPictureBG*{
  \AtPageUpperLeft{%
    \put(0,-40){\raisebox{15pt}{\makebox[\paperwidth]{\begin{minipage}{21cm}\centering
      \textcolor{gray}{This article has been accepted for publication in the proceedings of the \\2023 IEEE International Conference on Wireless and Mobile Computing, Networking And Communications (WiMob).\\DOI: 10.1109/WiMob58348.2023.10187799} 
    \end{minipage}}}}%
  }
  \AtPageLowerLeft{%
    \raisebox{25pt}{\makebox[\paperwidth]{\begin{minipage}{21cm}\centering
      \textcolor{gray}{ \copyright 2023 IEEE.  Personal use of this material is permitted.  Permission from IEEE must be obtained for all other uses, in any current or future media, including reprinting/republishing this material for advertising or promotional purposes, creating new collective works, for resale or redistribution to servers or lists, or reuse of any copyrighted component of this work in other works.
      }
    \end{minipage}}}%
  }
}

\begin{document}
\title{Latency and Power Consumption in 2.4\,GHz IoT Wireless Mesh Nodes: An Experimental Evaluation of Bluetooth Mesh and Wirepas Mesh}

\author{\IEEEauthorblockN{Silvano Cortesi\IEEEauthorrefmark{1}, Christian Vogt\IEEEauthorrefmark{1}, Elio Reinschmidt\IEEEauthorrefmark{1} and Michele Magno\IEEEauthorrefmark{1}}
\IEEEauthorblockA{\IEEEauthorrefmark{1}\textit{Department of Information Technology and Electrical Engineering, ETH Zurich, Zurich, Switzerland}} \IEEEauthorblockA{name.surname@pbl.ee.ethz.ch}
}

\maketitle
\begin{abstract}
The rapid growth of the \acl{IoT} paradigm is pushing the need to connect billions of battery-operated devices to the internet and among them. 
To address this need, the introduction of energy-efficient wireless mesh networks based on Bluetooth provides an effective solution. This paper proposes a testbed setup to accurately evaluate and compare the standard Bluetooth Mesh 5.0 and the emerging energy-efficient Wirepas protocol that promises better performance. The paper presents the evaluation in terms of power consumption, energy efficiency, and transmission latency which are the most crucial features, in a controlled and reproducible test setup consisting of 10 nodes. Experimental results demonstrated that Wirepas has a median latency of \SI{2.83}{\milli\second} in \acl{LL} mode respectively around \SI{2}{\second} in the \acl{LE} mode. The corresponding power consumption is \SI{6.2}{\milli\ampere} in \acl{LL} mode and \SI{38.9}{\micro\ampere} in \acl{LE} mode. For Bluetooth Mesh the median latency is \SI{4.54}{\milli\second} with a power consumption of \SI{6.2}{\milli\ampere} at \SI{3.3}{\volt}. Based on this comparison, conclusions about the advantages and disadvantages of both technologies can be drawn.

\end{abstract}
\acresetall
\vspace{-1em}
\section{Introduction}

The \ac{IoT} is growing at a rapid pace. It is estimated that there will be over 100 billion \ac{IoT} devices by 2025~\cite{iot_prediction}. Furthermore, more and more of today's \ac{IoT} devices are battery-operated and connected not only to the internet gateway but also to other neighboring devices to expand the network coverage~\cite{cilfone_wireless_2019}. The introduction of energy-efficient mesh networks can provide an effective solution~\cite{alma99118120421405503}, which could even be self-sustainable~\cite{zhou_self-sustainable_2019}. In a mesh network, the gateway is the only one that needs to be connected to the internet - the other nodes form the infrastructure by connecting directly, dynamically, and non-hierarchically to route data~\cite{mesh_general,alma99118120421405503}. A new node only needs to be within radio range of an existing node to become part of the network~\cite{baert_bluetooth_2018}, enabling a rapid roll-out of sensor nodes or even dynamic readout~\cite{trotta_bee-drones_2019}.

The IEEE 802.15.1 physical layer and on top the \ac{BLE} is the most popular technology for \ac{IoT} devices and energy-efficient communication~\cite{kindt2015adaptive}. It is predicted that 31\% of all \ac{IoT} device shipments will use \ac{BLE} for communication by 2024~\cite{ble_growth}, including Bluetooth Mesh~\cite{bluetooth_sig_mesh_2019, darroudi_bluetooth_2020, yin_survey_2019} enabled devices. Today, more and more application scenarios are increasing the interest on the mesh configuration with \ac{BLE}~\cite{leon_experimental_2020,leonardi_multi-hop_2018}. 

Bluetooth Mesh is an open mesh protocol, based on \ac{BLE} and IEEE 802.15.1. In most cases, the protocol uses advertisement packets sent in a flooding manner to transmit packets from a source to its destination~\cite{baert2018bluetooth}. Due to the underlying \ac{BLE} layer, it is limited to the three \ac{GAP} advertisement channels for data communication, and can therefore not compensate for congestion on these channels. Additionally, relay nodes need an always-on radio, as they constantly need to scan for advertisements/packets~\cite{baert2018bluetooth}.
In the recent year, many mesh solutions have been proposed for energy-efficiency using the same physical layer of \ac{BLE}~\cite{ray2019edge}. The majority of the previous work was proposed by academic research and very few are today commercial solutions.

Wirepas Mesh succeeded in boosting up from university to a commercial reality and today is a promising mesh protocol based on the same physical layer as \ac{BLE}~\cite{juopperi2019real}. It implements its own mesh-oriented protocol for data transfer and routing, using all \ac{BLE} channels together with \ac{FDMA} and \ac{TDMA}. Exploiting this architecture, Wirepas Mesh achieves to create an energy-efficient and scalable mesh network, allowing for a high device density~\cite{wp_scal}. Additionally, in contrast to Bluetooth Mesh, a part of the data traffic is sent via a routed protocol~\cite{perez2020bluetooth,wp_routing}. By dynamically adjusting the transmission power depending on the path loss to the next router, the transmission power is additionally optimized for the sake of energy efficiency~\cite{wp_intro}.

Previous literature presented already investigations into the power consumption of Bluetooth Mesh based on the popular commercial nRF51422 \ac{SoC} from Nordic Semiconductors~\cite{darroudi_bluetooth_2019}, but other promising and emerging mesh technologies such as Wirepas are not yet well covered. Moreover, often the evaluation and comparison lack a reproducible setup or are too strongly linked to the used setup. Whilst, next to the power consumption, the performance and delay in Bluetooth Mesh networks are well profiled by now \cite{darroudi_bluetooth_2020, yin_survey_2019}, other mesh networks such as Wirepas are not.

In this paper, we present a testbed design for real-world testing of mesh networks with controlled connections. Two different technologies, Bluetooth Mesh and the promising Wirepas Mesh are accurately evaluated and compared in terms of packet forwarding behavior (delay), power consumption, and energy-efficiency. This evaluation provides performance data to future sensor node designers, on which they can choose the best technology for their applications. Furthermore, the evaluation of the two mesh technologies is done in a fair and reproducible way to extract the most relevant features.  

The main contributions of this paper focus on three areas:
\begin{enumerate*}[,,font=\itshape]
\item design and implementation of a compact and scalable testbed for experiments with a controlled mesh topology based on the IEEE 802.15.1 physical layer and \SI{2.4}{\giga\hertz} nodes instead of the often used uncontrolled deployment in an environment, 
\item analysis and comparison of the two mesh networks in terms of latency as well as 
\item power consumption and energy used per node.
\end{enumerate*}

The paper first gives a background on the two observed mesh networks in section \ref{sec:background}, then the custom-developed testbed for a precisely defined network topology is explained in \ref{sec:materials}. After that, the results of latency and power measurements are presented in \ref{sec:results} before the paper is then concluded with \ref{sec:discussion}.

\section{Background on Mesh Networks} \label{sec:background}
    \subsection{Bluetooth Mesh}
        Bluetooth Mesh~\cite{bluetooth_sig_mesh_2019} is a Mesh technology with the goal to expand \ac{BLE} with many-to-many device communication over the \ac{BLE} radio - using the full \ac{BLE} protocol stack. To expand the range of the network, it uses the advertising/scanning roles of \ac{BLE} to implement managed flooding mechanism. Whenever a node sends a message, it broadcasts a packet over the advertising channels. Every node in the network, that hears the message, relays it for the other nodes in the network to hear. The simplicity of this Mesh technology, the low cost of implementation, and the wide use of \ac{BLE} make Bluetooth Mesh an attractive candidate for \ac{IoT} applications.
        \subsubsection{Controlled Flooding}
            Bluetooth Mesh uses no routing mechanism. Instead, controlled flooding is performed. When a node sends a message, it is sent to all available nodes within radio range. These messages are picked up by the other nodes and relayed until the message reaches its destination. This introduces a lot of redundant traffic, thus creating congestion whilst augmenting reliability by eliminating single point of failure. To optimize this, each node relays the same message only once and the message is equipped with a \ac{TTL}. Each node then decrements the \ac{TTL} until it is zero.
            \par
            In order for the network to maintain a list of active devices, each node sends a heartbeat message in regular time intervals specified by the user.
        \subsubsection{Node features}
            There are several features, also called roles or operating modes that can be assigned to a node. The for this work relevant ones are:
            \paragraph{Relay node}
                Re-transmits all received messages from the network. Due to the protocol being based on \ac{BLE} advertisements, Relay nodes need to always listen.
            \paragraph{Low-Power node}
                Spends most of the time in sleep, and wakes up periodically (or event-driven) to broadcast a message. In order to receive a message, he needs to poll a so-called friend node (can be a Relay node) for messages.
    \subsection{Wirepas Massive}
        Wirepas Massive is a proprietary, decentralized, and self-healing Mesh Solution of the Finnish company Wirepas. It offers scalability and reliability, while being energy efficient and easy to configure~\cite{wp_intro}. The Protocol is based on the same physical layer as \ac{BLE} and can thus be used with the same hardware as Bluetooth Mesh~\cite{wp_radio} - although the software framework is so far only ported to a few \acp{MCU}.
        \subsubsection{Network structure}
\begin{figure}[htpb!]
                \centering
                \includesvg[width=0.65\columnwidth]{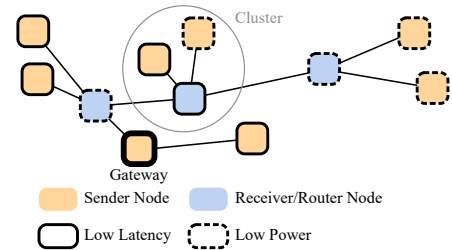}
                \caption{Wirepas Mesh: Network structure showing an interconnection of \acl{LE} and \acl{LL} nodes acting as sender and router.}
                \label{fig:wirepas_network_structure}
            \end{figure}

            The Wirepas Mesh consists of nodes and sinks/gateways, an example is given in Fig.~\ref{fig:wirepas_network_structure}. A node can either be a router or non router. By default, the role is selected autonomously. Nodes try to optimize their path to the sink node using the best router nodes. The nodes in Wirepas Mesh form together clusters and in each cluster, exactly one node is a router (also called cluster head). All other nodes in the cluster establish a one-to-one connection with the router and will be non-router nodes. The routers themselves connect to each other to link the clusters together.
            Additionally, there is the possibility to be a Non-Router Long Sleep node, which has to poll messages from its router and is thus not continuously connected to the network.
        \subsubsection{Messages and Routing}
            Each message in a Wirepas Mesh can be addressed to a single device (unicast), to all devices (broadcast), or to a group of devices (multicast). Thereby it can have one of three directions:
            \begin{itemize}
            \item\textbf{Uplink:} From node to sink
            \item\textbf{Downlink:} From sink to node(s)
            \item\textbf{Intra network:} From node to node(s)
            \end{itemize}
            Depending on the message direction and addressing there are two different routing mechanisms: Using \textbf{Adaptive Routing}, messages are sent using the mesh routing tree on a hop-by-hop basis with acknowledgment from and to the sink. For intra-network communication, \textbf{Adaptive Flooding} is applied.
        \subsubsection{Operation modes}
            Each Wirepas Mesh node can work in one of two modes: \textbf{\acf{LL}} mode and \textbf{\acf{LE}} mode. Devices with different modes can be mixed within the same Mesh network (see Fig.~\ref{fig:wirepas_network_structure}).
            \paragraph{\acl{LL} mode}
                The \ac{LL} mode is intended for high throughput and low latency applications. This is achieved by routers always listening during their idle time, allowing nodes to transmit immediately to the next hop.
            \paragraph{\acl{LE} mode}
                In the \ac{LE} mode, the node can achieve the lowest possible power consumption - including routers. This is achieved by the devices being active only during a short scheduled interval (2, 4, or 8 seconds period).
        \subsubsection{Physical Layer}
            Wirepas Mesh, using the same physical layer as \ac{BLE}, uses a pre-configured, fixed channel to do the provisioning - but uses then all \ac{BLE} channels for communication. This enables high-density networks by using a combination of \ac{TDMA} and \ac{FDMA} to minimize congestion and interference. In addition, the nodes adaptively change their transmission power to a minimum, such that only nodes within the clusters are within range.
\section{Materials and Methods}\label{sec:materials}
    \subsection{Materials}
        \subsubsection{Hardware}
            \begin{figure}[htpb!]
                \centering
                \includesvg[width=\columnwidth]{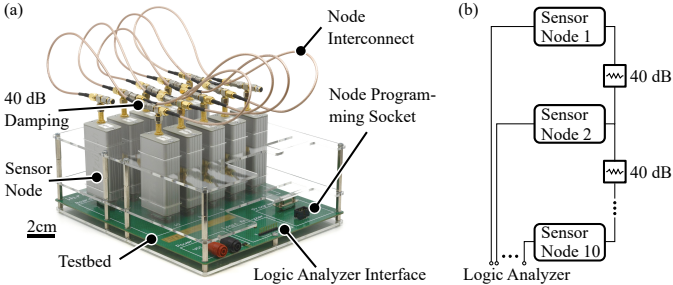}
                \caption{Hardware setup with 10 nodes in aluminium housing on a testbed for measurements and programming (a) and schematic connection overview of the nodes (b).}
                \label{fig:hardware}
            \end{figure}
            To provide a controlled evaluation, custom nodes were designed based on Würth Protheus III modules (employing nRF52840 \acsp{SoC}), as implementations for Bluetooth Mesh as well as Wirepas Mesh exist for these \acsp{SoC}. The nodes are connected with \SI{20}{\centi\meter} \acs{SMA} cables and \SI{40}{\deci\bel} attenuators between each node. This allows a pre-defined connection topology between the nodes, as a signal passing 2 nodes is attenuated by \SI{80}{\deci\bel}.  The nodes are enclosed in aluminum housings to reduce cross-talk between the nodes. All nodes are grouped on a testbed to facilitate measurements and programming (see Fig.~\ref{fig:hardware}a). Additionally, each node provides one pin wired to a pin header for latency evaluation (Fig.~\ref{fig:hardware}b). Pins used during measurements (outputs and VCC) are additionally filtered with \ac{EMI} filter providing \SI{-37}{\deci\bel} at \SI{2.4}{\giga\hertz} to avoid cross-talk of the nodes via the shared power supply when plugged into the testbed. More nodes could easily be added with a second testbed, and different network topologies exploited by re-wiring the SMA cables. The design of the custom nodes and the testbed are released as open-hardware on GitHub~\cite{github_repo}.
            
        \subsubsection{Firmware}
            The firmware for the nodes is written in Zephyr v3.1.99 for the Bluetooth Mesh evaluation and Wirepas Mesh SDK 1.3.0 with Stack 5.2.0.53 for the Wirepas Mesh evaluation. For Wirepas, both the \ac{LL} and \ac{LE} mode have been implemented. For Bluetooth Mesh, no Low-Power nodes are established (as they can not route traffic).
            
            Whenever a message is received, a callback in the the corresponding node's firmware toggles the output pin, enabling latency measurements by observing the pin toggling. On Wirepas, the interrupt callback is called after the messages have been relayed, on Bluetooth Mesh, it is called before the message is relayed. 
            
        \subsubsection{Experimental setup}
            \begin{figure}[htpb!]
                \centering
                \includesvg[width=0.8\columnwidth]{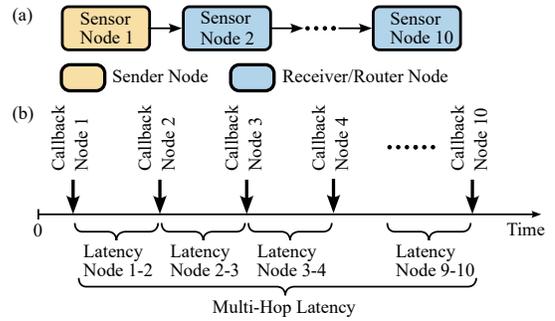}
                \caption{Network topology of the 10 nodes used in the experimental setup with one sender and 9 relaying nodes, all connected in a linear topology (a) and latency definition for measurements (b).}
                \label{fig:network_topology}
            \end{figure}
            The experimental setup consists of 10 nodes in a chain topology (see Fig.~\ref{fig:network_topology}a). This topology has been chosen in order to maximize the number of hops for a message in the test setup. The first node (Node 1) is the sender, transmitting messages of \SI{1}{\byte} length to the other nodes. All the other nodes (Node 2-10) are relays and receivers at the same time. As each node should communicate only with its neighbor (in order to have a known routing path), the transmission power of all nodes is set to \SI{-40}{\deci\belmilliwatt}. Together with the \SI{40}{\deci\bel} attenuators, the receiver sensitivity of \SI{-93}{\deci\belmilliwatt}, careful shielding of nodes, and testbed design, the  cross-talk between non-neighboring nodes is avoided.
            \par
            For each of the three modes (Wirepas \ac{LL}, Wirepas \ac{LE}, Bluetooth Mesh) a latency measurement, as well as a power measurement is performed. For the latency measurement, the first node sends a broadcast message to all other nodes and toggles then its GPIO pin. Whenever a routing node receives this message and toggles its GPIO, the logic analyzer attached to the pin header (DreamSourceLab DSLogic Plus) will register it. The difference between the GPIO toggling of two consecutive nodes is defined as latency (see Fig.~\ref{fig:network_topology}b). 
            
            For latency measurements, a total of 1000 measurement cycles (i.e. messages routed through the mesh) were recorded at a sampling frequency of \SI{1}{\mega\samplepersecond}.  All nodes are directly supplied by an external \SI{3.3}{\volt} power source unit (Keysight E36313A) during latency measurements.
            \par
            For the power consumption measurement, the setup remained the same, but the transmission power of the routing node currently measured was set to \SI{0}{\deci\belmilliwatt} to provide a more realistic setting. Using a sampling frequency of \SI{100}{\kilo\samplepersecond}, the Nordic Power Profiler Kit 2 was used to record a \SI{3}{\second} window, centered around a toggle of the output pin, indicating the reception and forwarding of the message through the node.
\section{Results}\label{sec:results}
    This section presents the measurements done to evaluate and compare the two target technologies. In particular, the latency and power consumption are accurately profiled up to 10 nodes. Energy-efficient considerations are then derived from those two main measurements. 
    
    \subsection{Latency}
    Latency measurements are split into two parts:
    \begin{enumerate*}[,,font=\itshape]
    \item the Bluetooth Mesh and Wirepas \ac{LL} mode and 
    \item the Wirepas \ac{LE} mode.
    \end{enumerate*}
    The parts are selected such, that at least similar setups are compared with each other.
        \subsubsection{Bluetooth Mesh and Wirepas \acf{LL}}
        An overview of the measured Bluetooth Mesh and Wirepas \ac{LL} is given in Tab.~\ref{tab:res_lat_ble_mesh_wpll}. The results are split into 1 and 9-hop configurations, with 9 hops being the maximum measurable distance in the experimental setup. 
        
        Bluetooth Mesh shows with an average of \SI{8.86}{\milli\second} and a median of \SI{4.54}{\milli\second} a larger delay than the comparable 1-hop Wirepas Mesh setup with \SI{5.21}{\milli\second} and \SI{2.83}{\milli\second} respectively. 
        
        A similar trend is visible in the 9-hop case, showing Bluetooth Mesh with larger average latency but a smaller standard deviation (\SI{35.13}{\milli\second} compared to \SI{121.88}{\milli\second} of Wirepas). The increased standard deviation of Wirepas is attributed mostly to latency outliers of up to \SI{1.78}{\second}. The largest measured outlier of Bluetooth Mesh was \SI{282.16}{\milli\second}.
        
            \begin{table}[htpb!]
                \centering
                \caption{Bluetooth Mesh and Wirepas \acl{LL} Latencies}
                \label{tab:res_lat_ble_mesh_wpll}
                \begin{tabular}{|r|rr|rr|}
                    \hline
                    \multirow{2}{*}{}  & \multicolumn{2}{c|}{\textbf{Bluetooth Mesh}} & \multicolumn{2}{c|}{\textbf{Wirepas Mesh LL}} \\ \cline{2-5} 
                                       & \multicolumn{1}{c|}{\textbf{1-Hop}}             & \multicolumn{1}{c|}{\textbf{9-Hop}} 
                                                & 
                                        \multicolumn{1}{c|}{\textbf{1-Hop}}
                                                &
                                         \multicolumn{1}{c|}{\textbf{9-Hop}}        \\ \hline
                    \textbf{Average}   & \multicolumn{1}{r|}{\SI{7.86}{\milli\second}}   & \SI{70.57}{\milli\second}  & \multicolumn{1}{r|}{\SI{5.21}{\milli\second}}  & \SI{50.28}{\milli\second} \\ \hline
                    \textbf{Median}    & \multicolumn{1}{r|}{\SI{4.54}{\milli\second}}   & \SI{63.44}{\milli\second}  & \multicolumn{1}{r|}{\SI{2.83}{\milli\second}}  & \SI{29.76}{\milli\second}\\ \hline
                    \textbf{Std. Dev.} & \multicolumn{1}{r|}{\SI{11.43}{\milli\second}}  & \SI{35.13}{\milli\second} & \multicolumn{1}{r|}{\SI{34.30}{\milli\second}} & \SI{121.88}{\milli\second} \\ \hline
                    \textbf{Minimum}   & \multicolumn{1}{r|}{\SI{3.17}{\milli\second}}   & \SI{37.94}{\milli\second}  & \multicolumn{1}{r|}{\SI{16}{\micro\second}}    & \SI{25.72}{\milli\second}\\ \hline
                    \textbf{Maximum}   & \multicolumn{1}{r|}{\SI{111.42}{\milli\second}} & \SI{282.16}{\milli\second}  & \multicolumn{1}{r|}{\SI{1.55}{\second}}        & \SI{1.78}{\second}\\ \hline
                \end{tabular}
            \end{table}

\begin{figure}
                \centering
                \includesvg[width=0.8\columnwidth]{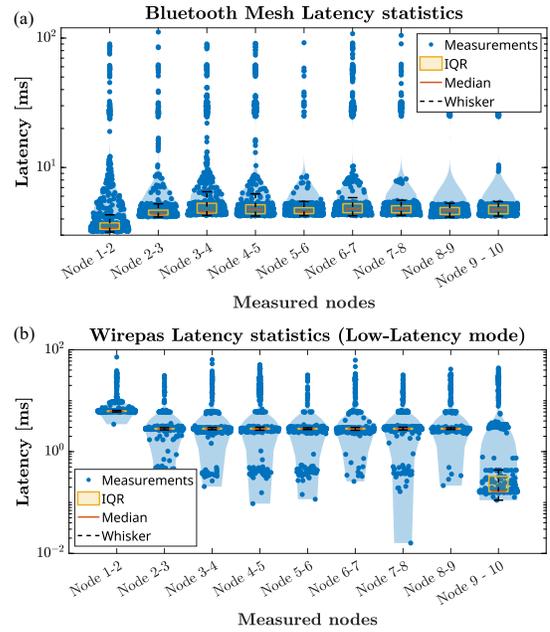}
                \caption{Overview of single hop latency measurements grouped according to node connections for both Bluetooth Mesh (a) as well as Wirepas \ac{LL} (b).}
                \label{fig:ble_wpllviolin}
            \end{figure}

            A plot over all measurements is given in Fig.~\ref{fig:ble_wpllviolin}. Each latency measurement is one blue dot, grouped by node connections for better understanding between the source node and the routing nodes. Whereas both protocols show outliers of at least \SI{100}{\milli\second}, most transmission delays are closely spaced, as indicated by the individual median values as well as \ac{IQR} including any Whisker. 
            
            Between Bluetooth Mesh (Fig.~\ref{fig:ble_wpllviolin}a) and Wirepas (Fig.~\ref{fig:ble_wpllviolin}b) a few differences are visible: 
            
            While the outliers of Bluetooth Mesh range up to \SI{111.42}{\milli\second} for single hop connections, no latencies below \SI{3.17}{\milli\second} were observed. In contrast, Wirepas latencies also range significantly below the median latency value (down to \SI{0.17}{\milli\second}).
            
            Bluetooth shows a significantly lower latency on Node 1-2 (median \SI{3.4}{\milli\second}) than on the rest of the connections (median \SI{4.4}{\milli\second}). This trend is opposite in the Wirepas test case, where Node 1-2 reports higher latency than the rest, with a median of \SI{6.25}{\milli\second} compared to \SI{2.8}{\milli\second}. This is attributed to a different point in time where the callback for the latency measurement is issued (at reception of packet or at transmission). A further indication is the last hop between Node 9-10 for Wirepas, having a median time of approx. \SI{0.17}{\milli\second}.
            
            Furthermore, the distribution of Bluetooth Mesh latencies is with approx. \SI{10}{\milli\second} wider than the Wirepas setup in Fig.~\ref{fig:ble_wpllviolin}. The individual Bluetooth Mesh connections are divided into 3 distinct delays with inter-delay distances of \SI{0.4}{\milli\second} between \SI{4.2}{\milli\second} and \SI{5.4}{\milli\second}. In contrast, the split into these distinctive 3 delay levels was not observed in Wirepas, although Wirepas shows larger discrete delay slots of approx. \SI{4.2}{\milli\second} with an overall median latency of \SI{2.83}{\milli\second}.
            
\begin{figure}
                \centering
                \includesvg[width=0.8\columnwidth]{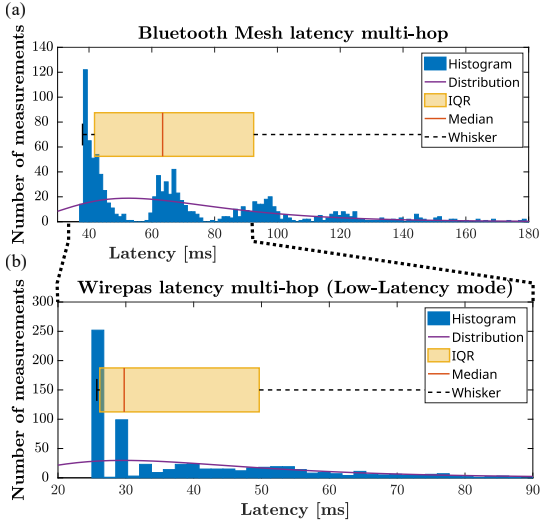}
                \caption{Multi-hop latency distribution for the Bluetooth Mesh setup (a) and the Wirepas setup (b). Longer latency outliers are omitted ($>$ 95\% of the measurement data is displayed within the figure).}
                \label{fig:ble_wpll_multi}
            \end{figure}

            Over all 9 hops, the Wirepas latencies (median \SI{29.76}{\milli\second}) are lower than the Bluetooth Mesh versions (median \SI{63.44}{\milli\second}, see Fig.~\ref{fig:ble_wpll_multi}). In contrast to Wirepas, Bluetooth Mesh displays regular latency intervals visible by the peaks in the histogram, indicating that some transmission delays are significantly longer than the regular transmission time, for example also depicted in Fig.~\ref{fig:ble_wpllviolin}a on the connection of Node 3-4 regularly showing delays of \SI{10}{\milli\second} or more compared to other node connections. Wirepas instead has an overall smooth histogram and 25\% of the data transmissions traveling through the 9 hops in minimum time (approx. \SI{28}{\milli\second}). The large outliers of the 9-hop Wirepas setup of up to \SI{1.78}{\second} are not visible in Fig.~\ref{fig:ble_wpll_multi}b to keep the graph readable (less than 10 events over the 1000 transmissions).

        \subsubsection{Wirepas Mesh - \acf{LE}}
        The by Wirepas offered \acl{LE} mode is designed to conserve more energy with the trade-off of reduced throughput and significantly increased latency between hops. Tab.~\ref{tab:res_lat_wp_lp} displays the statistics of 1000 messages passed through such a mesh. As median, the message relaying required \SI{1.26}{\second} between two nodes, with a maximum delay of up to \SI{7.5}{\second} for a single hop and \SI{26.86}{\second} for all 9 hops.
        
            \begin{table}[htpb!]
                \centering
                \caption{Wirepas Mesh Latency, \acl{LE} mode}
                \label{tab:res_lat_wp_lp}
                \begin{tabular}{|r|rr|}
                    \hline
                    \multirow{2}{*}{}  & \multicolumn{2}{c|}{\textbf{Wirepas Mesh LE}} \\ \cline{2-3} 
                                       & \multicolumn{1}{c|}{\textbf{1-Hop}}      & \multicolumn{1}{c|}{\textbf{9-Hop}} \\ \hline
                    \textbf{Average}   & \multicolumn{1}{r|}{\SI{1.41}{\second}}  & \SI{12.72}{\second} \\ \hline
                    \textbf{Median}    & \multicolumn{1}{r|}{\SI{1.26}{\second}}  & \SI{12.87}{\second} \\ \hline
                    \textbf{Std. Dev.} & \multicolumn{1}{r|}{\SI{0.74}{\second}}  & \SI{2.13}{\second} \\ \hline
                    \textbf{Minimum}   & \multicolumn{1}{r|}{\SI{0.48}{\second}}  & \SI{10.85}{\second} \\ \hline
                    \textbf{Maximum}   & \multicolumn{1}{r|}{\SI{7.50}{\second}}  & \SI{26.86}{\second} \\ \hline
                \end{tabular}
            \end{table}
            
\begin{figure}
                \centering
                \includesvg[width=0.8\columnwidth]{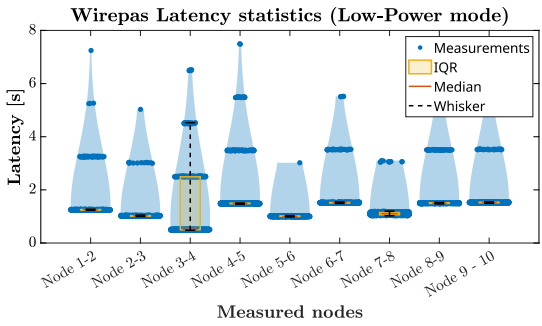}
                \caption{Wirepas Mesh \acl{LE} mode: Violin plot of the single hops}
                \label{fig:wp_lp_violin}
            \end{figure}

        The set 2-second routing interval is clearly visible in the latency distribution of Fig.\ref{fig:wp_lp_violin}. Among the node connections, the latencies are shifted but sum up to \SI{2}{\second} for 2 consecutive connections.
            
\begin{figure}
                \centering
                \includesvg[width=0.8\columnwidth]{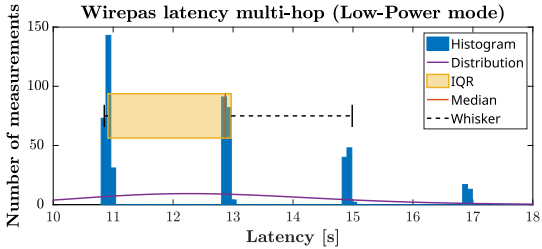}
                \caption{Wirepas Mesh \acl{LE} mode: Multi-hop Latency distribution}
                \label{fig:wp_lp_multi}
            \end{figure}

        Similarly, the multi-hop delay histogram in Fig.~\ref{fig:wp_lp_multi} clearly indicates the 2-second intervals with the latencies located around multiples of 2 seconds. Notably, the forwarding sometimes misses a transmission interval, causing a larger spread over time. In comparison, less than 150 of the 1000 messages arrived within the minimum transmission time for Wirepas \ac{LE}. However, for Wirepas \ac{LL} over 250 arrived on time (factor 1.67).
            
    \subsection{Power and energy consumption}
    The power consumption is measured individually for the nodes during their normal operation within the network.
    
    Given the overview in Tab.~\ref{tab:res_power}, neither Bluetooth Mesh nor Wirepas \acl{LL} show a significant difference in power consumption, at an average power consumption of \SI{20.5}{\milli\watt}. In contrast, Wirepas in \ac{LE} mode offers significant energy savings (\SI{128.34}{\micro\watt}, a factor of approx. 160 to \ac{LL} mode) at the cost of significantly higher latency (\SI{2.83}{\milli\second} compared to \SI{1.26}{\second} per hop, a factor of approx. 445).
        \begin{table}[htpb!]
            \centering
            \caption{Power Consumption of the Mesh Nodes}
            \label{tab:res_power}
            \begin{tabular}{|l|r|r|r|}
                \hline
                                         & \multicolumn{1}{c|}{\textbf{Idle}} & \multicolumn{1}{c|}{\textbf{Relay}} & \multicolumn{1}{c|}{\textbf{Average}} \\ \hline
                \textbf{Bluetooth Mesh}  & \SI{20.6}{\milli\watt} & \SI{19.83}{\milli\watt} & \SI{20.43}{\milli\watt}\\ \hline
                \textbf{Wirepas Mesh LL} & \SI{20.61}{\milli\watt} & \SI{20.88}{\milli\watt} & \SI{20.53}{\milli\watt}\\ \hline
                \textbf{Wirepas Mesh LE} & \SI{11.81}{\micro\watt} & \SI{3.03}{\milli\watt} & \SI{128.34}{\micro\watt}\\ \hline
            \end{tabular}
        \end{table}
        
\begin{figure}
            \centering
            \includesvg[width=0.78\columnwidth]{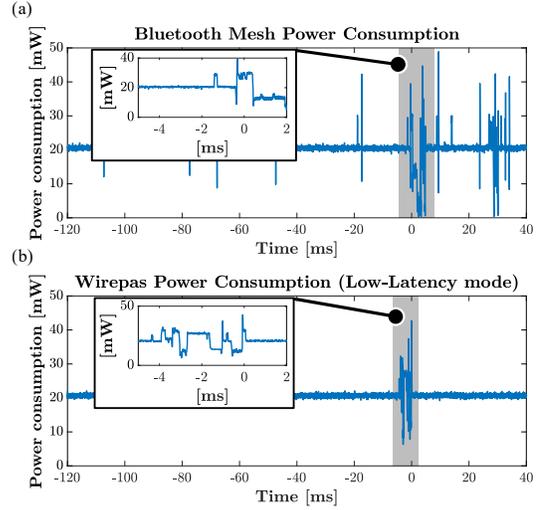}
            \caption{Bluetooth Mesh power consumption (a) and Wirepas Mesh \acs{LL} (b) with the callback event at \SI{0}{\second} and the zoomed in region marked in gray.}
            \label{fig:ble_wpll_pwr}
        \end{figure}

\begin{figure}
            \centering
            \includesvg[width=0.78\columnwidth]{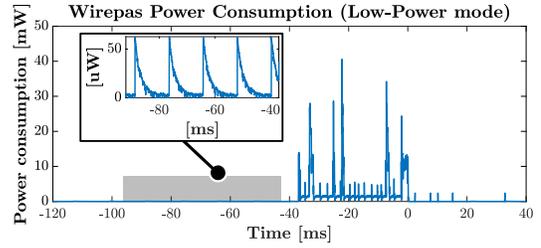}
            \caption{Wirepas Mesh \acs{LE} power consumption with the callback event at \SI{0}{\second} and the zoomed in region marked in gray.}
            \label{fig:wp_lp_pwr}
        \end{figure}

        A more detailed view of the power requirement during mesh operation is given in Fig.~\ref{fig:ble_wpll_pwr} to Fig.~\ref{fig:wp_lp_pwr}. The differences in the power consumption are visible in Fig.~\ref{fig:ble_wpll_pwr}a and Fig.~\ref{fig:ble_wpll_pwr}b within approx. \SI{5}{\milli\second} before and after the packet forwarding. Differences are attributed to different time points when the callback recorded is issued as well as the different internal design of the protocol. During the rest of the measurement, the node power consumption stays constant, as they are in receive mode only waiting for new packets to forward.
        
        In contrast, the Wirepas \ac{LE} mode depicted in Fig.~\ref{fig:wp_lp_pwr} allows the \acs{SoC} to go to sleep (\SI{10}{\micro\watt}) between the scheduled packet arrival times. Furthermore, the sensor node wakes only for approx. \SI{40}{\milli\second}, further reducing the energy consumed down to \SI{121.2}{\micro\joule} per wake-up event.
        
\begin{figure}
                \centering
                \includesvg[width=0.78\columnwidth]{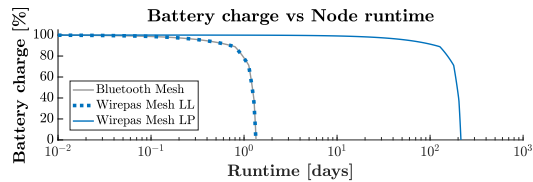}
                \caption{Estimated battery charge over node lifetime assuming a CR2032 cell with \SI{220}{\milli\ampere\hour} and an operating voltage of the nodes down to \SI{2.0}{\volt}}
                \label{fig:cap_estimate}
        \end{figure}

        In Fig.~\ref{fig:cap_estimate} the energy drain of the mesh configurations is simulated based on a node with a CR2032 coin cell battery (\SI{3}{\volt} nominal voltage, capacity of \SI{220}{\milli\ampere\hour}). A voltage of \SI{2.1}{\volt} is defined as empty battery. Due to the radio being always on for the Wirepas \ac{LL} nodes as well as the Bluetooth Mesh routing nodes, the battery is empty after approx. 1.4 days of operation. In contrast, the Wirepas \ac{LE} (routing) nodes could operate up to 210 days on this coin cell.
        
\section{Discussion and Conclusion}\label{sec:discussion}

Out of the mesh networks evaluated on an nRF52840, Bluetooth Mesh as well as Wirepas \ac{LL} show nearly identical power consumption of \SI{20.5}{\milli\watt} on average per node. Next to the power consumption, the latency and the distribution of latency is significantly different, with Wirepas offering on average a lower latency by a factor of approx. 1.5 and an overall lower latency spread especially for the multi-hop scenario. In contrast, Bluetooth Mesh also shows significant delay spread during single and multi-hop scenarios.

Despite the lower overall delay spread of Wirepas, a few significantly larger latencies of up to \SI{1.78}{\second} for the multi-hop case were measured within the Wirepas Mesh, whereas the Bluetooth Mesh only reached a maximum of \SI{282}{\milli\second}.

For ultra-low power applications, the preferred network would be the Wirepas \ac{LE} variant, consuming only \SI{121.2}{\micro\joule} (\SI{128.3}{\micro\watt} on average) per node and transmission, at the drawback of increased latencies by a factor of over 445. Such a node could run on a CR2032 coin cell for over 7 months.

This paper designed and presented an accurate open-hardware testbed for mesh networks based on the IEEE 802.15.1 physical layer, which is released on GitHub~\cite{github_repo}. In particular, the paper focused on the evaluations of the standard Bluetooth Mesh and the promising industrial standard Wirepas Mesh compared on the commercial popular Nordic nRF52840 as radio module. The performance has been presented in terms of latency and power and then discussed in terms of energy and overall performance.

\section*{Acknowledgments}
The authors want to thank Dr. Armin Wellig (Honeywell) and Honeywell Switzerland for their support, which made this work possible.

\bibliographystyle{IEEEtran}
\bibliography{bib/IEEEabrv, bib/references}
\end{document}